\documentclass[a4paper, pre, showkeys, preprintnumbers,amsmath,amssymb, preprint]{revtex4}

\usepackage[]{graphicx}
\usepackage{dcolumn}
\usepackage{bm}
\usepackage{amsmath, amssymb}
\begin{document}

\preprint{KU, NBI / Optical Tweezers Group}
\title{Allan Variance Analysis as Useful Tool\\to Determine Noise in Various Single-Molecule Setups}
\author{Fabian Czerwinski}
  \email{czerwinski@nbi.dk}
\author{Andrew C. Richardson}
\author{Christine Selhuber-Unkel}
\author{Lene B. Oddershede}
\affiliation{Niels Bohr Institute, University of Copenhagen\\
Blegdamsvej 17, 2100 K\o benhavn \O, Denmark}
\date{June 30, 2009}

\begin{abstract}
One limitation on the performance of optical traps is the noise inherently present in every setup. Therefore, it is the desire of most experimentalists to minimize and possibly eliminate noise from their optical trapping experiments. A step in this direction is to quantify the actual noise in the system and to evaluate how much each particular component contributes to the overall noise. For this purpose we present Allan variance analysis as a straightforward method. In particular, it allows for judging the impact of drift which gives rise to low-frequency noise, which is extremely difficult to pinpoint by other methods. We show how to determine the optimal sampling time for calibration, the optimal number of data points for a desired experiment, and we provide measurements of how much accuracy is gained by acquiring additional data points. Allan variances of both micrometer-sized spheres and asymmetric nanometer-sized rods are considered.
\end{abstract}


\keywords{optical tweezers, optical trapping, noise, drift, calibration, Allan variance, piezo stage, nanorods}

\maketitle

\section*{INTRODUCTION}
\label{sec:intro}  
Optical tweezers are the perfect nano-tool for single-molecule manipulation and investigations~\cite{Moffitt2008}.
With a correctly chosen wavelength they are nearly non-invasive and can be used to manipulate entire living 
microorganisms~\cite{Ashkin1987,Rasmussen2008} or
track organelles inside a cell~\cite{Tolic2004}.
 For single-molecule
investigations, often a handle, for example in the form of a dielectric micron-sized object, is attached to the
molecule of interest.
Optical tweezers can then follow the motion of the handle and calculate the forces exerted on the handle with sub-piconewton resolution.
The handle is often a polystyrene sphere with a diameter of a couple of micrometers,
but even nanoparticles such as gold spheres~\cite{Hansen2005}, gold nanorods~\cite{SelhuberUnkel2008},
 spherical silver nanoparticles~\cite{Bosanac2008}, or even individual quantum dots~\cite{Jauffred2008} can
be individually trapped and used as force transducers.

For single-molecule experiments it is extremely important to measure the 
distances moved and forces exerted by the single molecule as accurately as possible~\cite{Abbondanzieri2005}.
Therefore, it is crucial to minimize or eliminate noise and
drift and to perform force calibration as accurately as possible. Much effort has been put into 
minimizing noise, for instance entire setups have been covered to eliminate pressure fluctuations; the equipment is most
often placed on an optical table and sometimes even on a foundation which is separated from the rest of the
building. A successful way to eliminate drift is by using a laser beam parallel to the trapping laser to
 track the motion of a feducial marker which is 
attached in proximity to the handle of interest and thus subject to a similar drift~\cite{Carter2009}.
This setup, however, requires the use of at least two laser beams and two independent detection systems.
Another way to reduce drift that is often employed in optical trapping setups is to move the system of interest
away from any surface subject to significant drift. This could be done by using a dual-trap setup and
suspending the molecule of interest between two individual handles. But how efficient are these
methods? And which types of noise should one really worry about in optical trapping experiments? To answer
these questions it is essential to be able to quantify the noise introduced by each part of the equipment
or surroundings. 

Fourier analysis is an excellent tool to calibrate optical tweezers on-the-fly. Furthermore, peaks occuring in the high frequency part of the power spectrum can often easily be traced backwards to a particular
noise contribution. However, Fourier analysis is
not optimal for identifying and quantifying low-frequency noise, which drift, as inherently present in
experiments, produces. However, Fourier analysis has been used to judge the noise stemming
from particular experimental settings~\cite{Gittes1998,Klein2007}, but  for such an analysis
various assumptions have to be made about the bandwidth of the integration. Also, the regular positional
variance is often used as a measure for noise. However, the normal variance does not converge for purely stochastic
types of noise, such as white noise. 

In this Proceeding, we propose Allan variance analysis as a simple and efficient tool to pinpoint and quantify noise
in optical trapping facilities. The Allan variance of the positions visited by an optically trapped particle
can be calculated on-the-fly during an experiment and used, e.g., to determine the optimal length of 
a time series for accurate calibration~\cite{Gibson2008,Czerwinski2009}. By comparison to
simulated data with no drift present it has been explicitly shown that Allan variance is an excellent
tool to pinpoint and quantify low-frequency drift~\cite{Czerwinski2009, Czerwinski2009a}. Allan variance analysis has been specifically
used to verify that a CMOS camera had sufficient time resolution to track an optically trapped particle~\cite{Gibson2008}. In addition, Allan variance analysis is able to reveal noise contributions from commonly
used photodiode-based detection systems in optical trapping experiments~\cite{Czerwinski2009}, thus 
providing a platform for a qualified choice between, for example, a position-sensitive rather than a conventional quadrant photodiode for certain sets of experiments. Moreover, Allan variance has
revealed the impact of the piezo stage, the acoustic noise in the laboratory, and the geometry and stability 
of the sample chamber on the noise spectrum. Complementary and additional to Fourier analysis, Allan variance analysis
provides a basis for an optimal setup design~\cite{Czerwinski2009}. In this Proceeding, we review some of 
these results. In addition, we provide information about how the Allan variance of a measurement can be improved by
acquiring additional data points and about the Allan variance of an asymmetric gold nanorod. 

\section*{METHODS}
   \begin{figure}
   \begin{center}
   \begin{tabular}{c}
   \includegraphics[height=7cm]{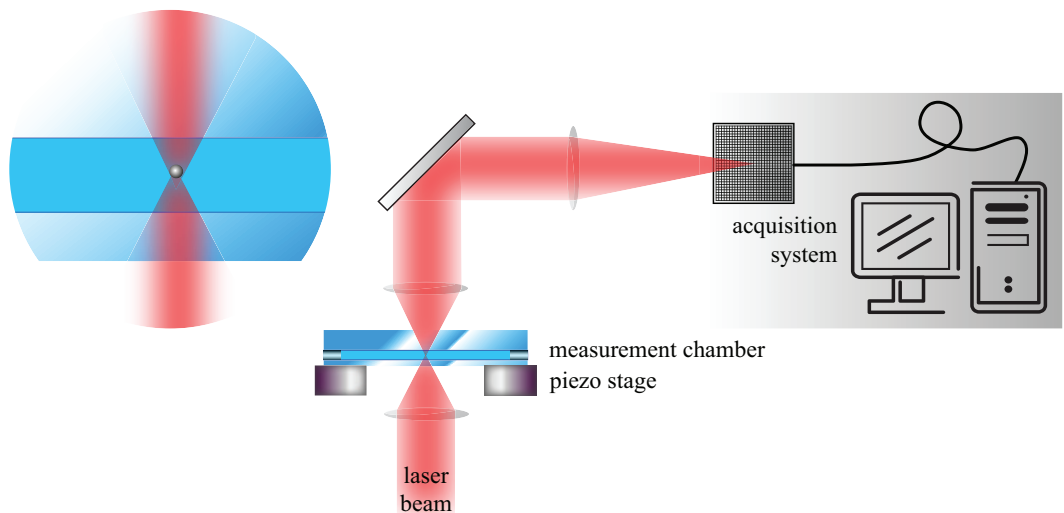}
   \end{tabular}
   \end{center}
   \caption{ \label{fig:sketch} 
Schematic drawing of experiments. A highly focused infrared laser beam traps a polystyrene sphere inside a custom-made measurement chamber that is mounted on a piezo stage. The forward scattered light is collected onto a photodiode that transmits the sphere's position into an acquisition system.}
   \end{figure} 
By focusing a laser beam into a measurement chamber that is mounted onto a piezo stage inside an inverted microscope (Leica DMIRBE), the optical trap is created. For trapping polystyrene speres (Bangs Laboratories, diameter $\left(800\pm10\right)$~nm) we used a water immersion objective (Leica, HCX, 63x, NA=1.20) at a color correction at its lowest setting (0.13~mm) and a measurement chamber of two cover slips (bottom: thickness 0.13--0.16~mm, top: 1~mm) sandwiched together by double sticky tape. The actual trap was formed in the middle of the chamber (total height $\left(95\pm5\right)~\mu$m).
For trapping gold nanorods~\cite{SelhuberUnkel2008}, an oil immersion objective (Leica, HCX PL Apo, 63?, NA =1.32) was used and the immersion oil was chosen to compensate spherical abberations~\cite{Reihani2007}. Here, the measurement chamber was custom-made from two cover slips (bottom: thickness 0.16--0.19~mm, top: 1~mm) held together by parafilm. The nanorods were trapped $5~\mu$m above the bottom. Their dimensions were for the longer axis $x=\left(63.8\pm7.4\right)$~nm and for the shorter axis $y=\left(37.3\pm5.0\right)$~nm (verified by transmission electron microscopy prior to measurements). In all cases, measurement chambers were sealed with vacuum grease to prevent evaporation. To ensure the optimal pointing stability of the laser beam, we switched on the laser at least one hour prior to experiments. All measurements were done at room temperature.

A schematic drawing of the essential parts of the experiment is shown in Figure~\ref{fig:sketch}. The forward scattered light is focused onto a position-sensitive photodiode (Pacific Silicon Sensor, DL100-7PCBA3) or onto a quadrant photodiode (Hamamatsu, S5981). Its output in voltage is connected through an amplifier, a low-pass filter of 100~kHz, and an acquisition card (National Instruments, PCI-6251) to a computer. By utilizing our data-streaming software that was custom-made in \textsc{Labview}~\cite{Czerwinski2009b}, we were not limited by the amount of acquired data (tested 2~h at 1~kHz, and 22~min at 100~kHz). 

Typically, for trapped spheres, the analysis was applied to adjacent time series of $2^{24}$ positions acquired at various acquisition frequencies. For nanorods, the time series consisted of $2^{21}$ positions acquired at 22~kHz. All additional filtering was carried out posterior to acquisition. For visual observation of the spheres right before and after experiments, a CCD camera (Sony, XC-ES50, 25~Hz) was used, whereas the nanorods could not be visualized optically.

\textbf{Calibration.} An optically trapped object experiences a harmonic force $F=-\kappa x$ with the trap stiffness $\kappa$ and the distance $x$ from the equilibrium position. Thus, $\kappa$ characterizes the thermal motion of the trapped object. For the analysis, the Langevin equation is typically solved and Fourier transformed. The result is a positional power spectrum that allows for finding the ratio between $\kappa$ and the friction coefficient $\gamma$, i.e. the corner frequency $f_c$:
\begin{equation} \label{eq:cornerfreq}
f_c = \frac{\kappa}{2 \pi \gamma}.
\end{equation}
In case a sphere is trapped far away from any surface, the Stokes law gives: $\gamma=6\pi r \eta$, where $r$ is the radius of the sphere and $\eta$ the viscosity of the surrounding medium; here, in all experiments, deionized water, $\eta=8.9\cdot10^{-4}$. The nanorods have the shape of sphero-cylinders. We approximated their shape as cylinders in order to calculate their drag coefficients accordingly.~\cite{SelhuberUnkel2008} The conversion factor $\beta$, which relates the distance measured in volts to the distance travelled in meters by the trapped object, was found by comparing theoretical and experimental diffusion constants~\cite{Oddershede2001}.

In a first step, we calculated the Allan variance to obtain the optimal measurement time for calibration. Secondly, we calibrated conditionally independent intervals of the time series with that particular length. We used the power-spectrum method as described previously~\cite{BergSorensen2004} with the freely available program~\cite{Hansen2006b}. Finally, we calculated various types of variances.

For typical noise phenomena found in nature, classical variances do not converge. For example, purely stochastic noise does not converge for the \textbf{normal variance}:
\begin{equation}\label{eq:variance}
\sigma^2\left(\tau\right) = \frac{1}{2}\left< \left(x_{i}-\bar{x}\right)^2 \right>_\tau\textnormal{,}
\end{equation}
where $\bar{x}$ denotes the mean of the time series and $x_{i}$ the mean of positions within the interval of length $\tau$.

The \textbf{Allan variance} is designed to converge for most naturally occuring noise.~\cite{Allan1966} Given a time series consisting of $N$ elements and a total measurement time of $t_\textnormal{acq}=f_\textnormal{acq}N$, the Allan variance is defined as: 
\begin{equation}	\label{eq:Allandefinition}
\sigma_x^2\left(\tau\right)=\frac{1}{2}\left< \left(x_{i+1}-x_i\right)^2 \right>_\tau\textnormal{,}
\end{equation}
where $x_i$ is the mean of the measurement interval $\tau$. In words, the Allan variance is half of the mean of the squared differences of neighboring intervals of a given length.

One can trade the Allan variance's conditional independence of neighboring intervals to gain a much smaller statistical error. For the \textbf{overlapping Allan variance}, one simply calculates all possible differences of neighboring intervals in a given time series. For a more comprehensive discussion we refer elsewhere~\cite{Czerwinski2009}.

As \textbf{thermal limit} for an object in an spatially confined trap, the standard error of an object's position averaged over the time interval $\tau$ is~\cite{Czerwinski2009}:
\begin{equation}
SE_{\left< x \right>}=\frac{1}{\sqrt{n}}\sqrt{\left< x^2 \right>} \approx 
\sqrt{ \frac{2 k_BT\gamma} {\kappa^2\tau} }\textnormal{,} \label{eq:thlimitonetrap}
\end{equation}
with $k_B T$ being the thermal energy. This limit cannot be bettered by any measurement of an object trapped by a single beam. Nevertheless, for so-called dual-beam traps this limit is about a factor 1.19 smaller~\cite{Gibson2008}.

Allan variances were calculated with a custom-made \textsc{Matlab} program~\cite{Czerwinski2008}. We measured Allan variances for various objects and parameters (trap stiffness, acquisition frequency, piezo) by acquiring three time series. Then the parameter was altered and we repeated the measurements. If not stated otherwise, we plotted the overlapping Allan variance in a log-log plot. All results stated in the following section are explicit and reproducible, although we chose to plot only particular data sets in order to keep a clear representation. 

\section*{RESULTS AND DISCUSSION}
   \begin{figure}
   \begin{center}
   \begin{tabular}{c}
   \includegraphics[width=10cm]{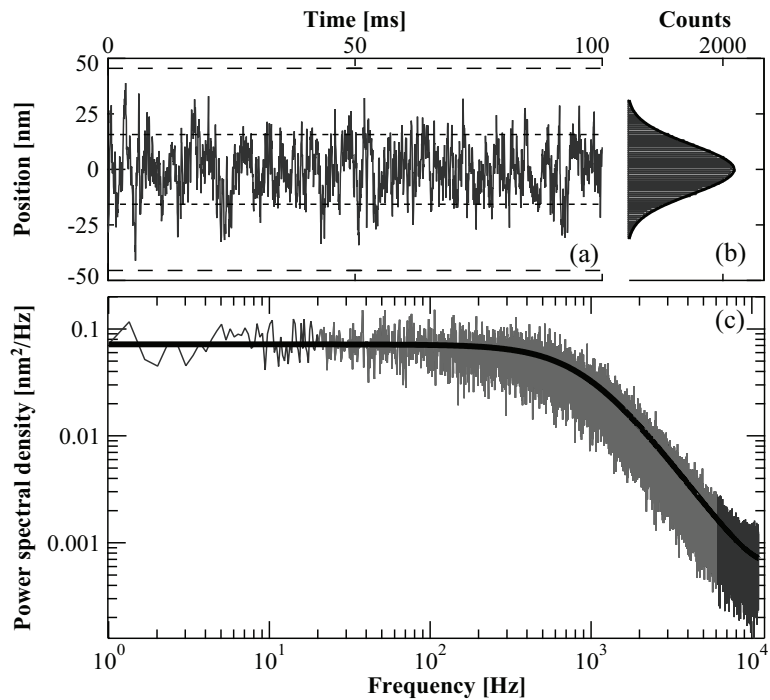}
   \end{tabular}
   \end{center}
   \caption[example] 
   { \label{fig:timeseries} 
Time series analysis of an optically trapped polystyrene sphere, $\kappa=67.7~\mu$m/pN. (a) Position as a function of time.
(b) Histogram of positions, fitted by a Gaussian distribution. (c) Power spectrum of positions. The full line
is a fit to the grey region which incorporated aliasing and filtering effects.}
   \end{figure} 
In the majority of the experiments a polystyrene sphere was trapped in an aqueous environment in the center of the measurement chamber such that the distance
to any surface was significantly larger than the diameter of the sphere. Using the position
sensitive diode the time series of the sphere was recorded and used for a calibration
routine which returned the corner frequency $f_c$, the trap stiffness $\kappa$, and 
the conversion factor $\beta$. Figure~\ref{fig:timeseries}~(a) shows
the time series of the positions visited by a trapped particle. The histogram of all positions visited is plotted in~(b), and is very well-fitted by
a Gaussian distribution. Figure~\ref{fig:timeseries}~(c) shows the corresponding power spectrum. Here, the light grey part is fitted by a Lorentzian function (black full line) which takes into account the filtering
effect of the photodiode~\cite{BergSorensen2003} and aliasing~\cite{BergSorensen2004} using programs described above~\cite{Hansen2006b}. Though the system is subject to low-frequency drift, this does not  show neither in the
time series, nor in the histogram, nor in the power spectrum, not even when compared to a simulation
of the situation without drift using similar
physical parameters~\cite{Czerwinski2009}.

\begin{figure}
   \begin{center}
   \begin{tabular}{c}
   \includegraphics[width=10cm]{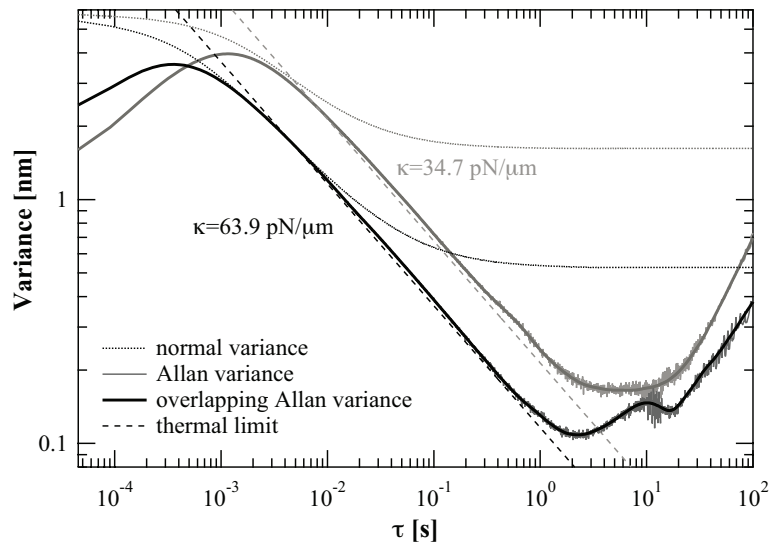}
   \end{tabular}
   \end{center}
   \caption[example] 
   { \label{fig:Allan}Variances of a polystyrene sphere trapped strongly, $\kappa$=63.9~pN/$\mu$m (black), or weakly, $\kappa$=34.7~pN/$\mu$m (grey).  The lighter graphs denote the Allan variances, the thicker ones the overlapping Allan variances.
Dashed lines are the thermal limits, dotted lines the normal variances.}
   \end{figure}

From time series as in Figure~\ref{fig:timeseries} the Allan variance can be calculated using Equation~(\ref{eq:Allandefinition})~\cite{Czerwinski2008}. A typical result is shown in Figure~\ref{fig:Allan} where the Allan variance is plotted as a function
of data acquisition time for a strong, $\kappa$=63.9 pN/$\mu$m (black), and a weak, $\kappa$=34.7 pN/$\mu$m (grey),
optical trap. The thicker lines denote the overlapping Allan variance for the same time series. The two dashed lines with
slopes of $-1/2$ correspond to the thermal limits (Equation~\ref{eq:thlimitonetrap}). For short measurement intervals the Allan variance
is smaller than the thermal limit, this is due to the correlation of the data points. The maximum of the 
Allan variance is at $\pi\tau_c$ where 
\begin{equation}
\pi \tau_c = \frac{1}{2 f_c}.
\end{equation}
For measurement intervals where the Allan variance is 
larger than the thermal limit, the data points are not correlated. The observation that the Allan variance is
very close to the thermal limit in the measurement time span between 100~ms and 1~s
is evidence of an extremely stable setup. The Allan variance has a global minimum at around 1--10~s, the
exact postion of this minimum is dependent on the trap stiffness. This  global minimum denotes the optimal
measurement time for calibrations. This minimum is the time where the Gaussian distributed parameters
have been measured for a time long enough to allow for their fairly precise determination while
the time interval is still short enough that drift is not yet a significant problem. Noticeable also, is the fact that
the stronger trap has a lower Allan variance, hence, a better accuracy than the weaker trap.
The dotted lines in Figure~\ref{fig:Allan} denote the normal variance (Equation~\ref{eq:variance}). The normal variance does not converge and
consequently does not provide the necessary resolution to determine the optimal measurement time.

\begin{figure}
   \begin{center}
   \begin{tabular}{c}
   \includegraphics[width=12cm]{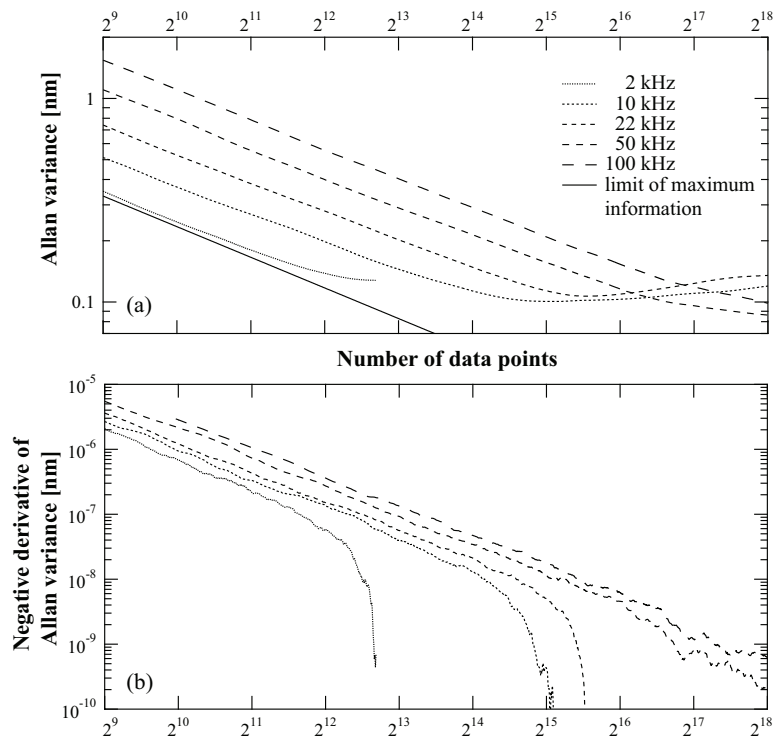}
   \end{tabular}
   \end{center}
   \caption[example] 
   { \label{fig:frequency} 
Frequency dependence of Allan variance and its negative derivative for a sphere trapped with $\kappa$=67.7~pN/$\mu$m. (a) Allan variance as a function of number 
of acquired data points. The full line denotes the limit of how much positional information a single data point could possibly hold.
(b) Negative derivative of Allan variance with respect to the number of data points.}
   \end{figure}

Figure~\ref{fig:frequency}~(a) shows the Allan variances as a function of the number of acquired data points. A polystyrene sphere is trapped with $\kappa$=67.7 pN/$\mu$m while the acquisition frequency is parameterized. The tested
frequencies range from 10~Hz to 100~kHz. For all frequencies, the acquisition of additional data points does not increase the accuracy above a certain threshold,
for instance at 2$^{15}$ for 10~kHz, because the Allan variance shows an absolute minimum. Furthermore, the closer a graph stays to limit of maximum information per individual data point, i.e. quasi the `thermal limit' in this way of representation, the more valuable is the acquisition of an additional data point.
Figure~\ref{fig:frequency}~(b) shows the negative values of the derivatives of the Allan variances with respect to the number of
data points -- it gives the increase in accuracy by acquiring more data points.  For a specific acquisition frequency, it passes through zero when its sampling has been optimal and for more points drift would start to dominate. 
Consistently, these graphs show that if the sampling frequency is 10~kHz, essentially nothing is
gained by acquiring more than 2$^{15}$ data points.

\begin{figure}
   \begin{center}
   \begin{tabular}{c}
   \includegraphics[width=10cm]{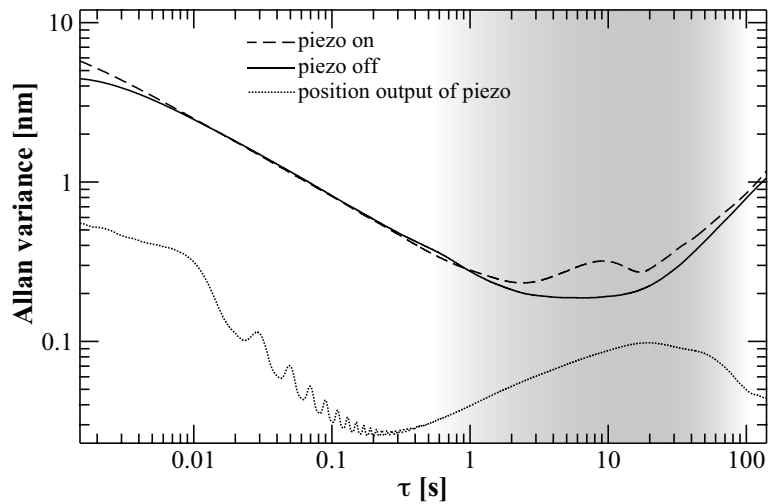}
   \end{tabular}
   \end{center}
   \caption[example] 
   { \label{fig:piezo} 
Contribution from the piezo stage to the Allan variance. The full line shows the Allan variance
of a trapped sphere when the piezo stage is off, the dashed line is the Allan variance
when the piezo is switched on. The dotted line is the Allan variance of the position output
from the piezo control box with a significant noise contribution over a broad frequency band indicated by grey shading.}
   \end{figure}

Often, the measurement chamber of an optical trapping experiment is mounted on a piezo-electric stage.
To investigate the possible contribution from the piezo stage to the noise spectrum we measured
the Allan variance for two similar trapping experiments. In Figure~\ref{fig:piezo} one sees the first case with the piezo switched off (full black line), and the second case with the piezo switched on (dashed line).
There is a distinct peak in the Allan variance at around 10~s which seems to originate from the
piezo stage. The lower dotted line in Figure~\ref{fig:piezo} is the Allan variance of the position of the
piezo stage as given by the piezo control box. This noise spectrum also peaks at around 10~s and
hence supports the conclusion that the piezo stage itself contributes to the noise spectrum over a broader low-frequency band as indicated by the grey shading. At short
measurement times the Allan variance of the piezo output has some oscillations which correspond to the
odd-numbered divisors of the piezo's resonance frequency of 100~Hz. The fact that the piezo contributes to the
noise spectrum implies that if it is not strictly needed for a particular experiment, it should be switched off.

\begin{figure}
   \begin{center}
   \begin{tabular}{c}
   \includegraphics[width=10cm]{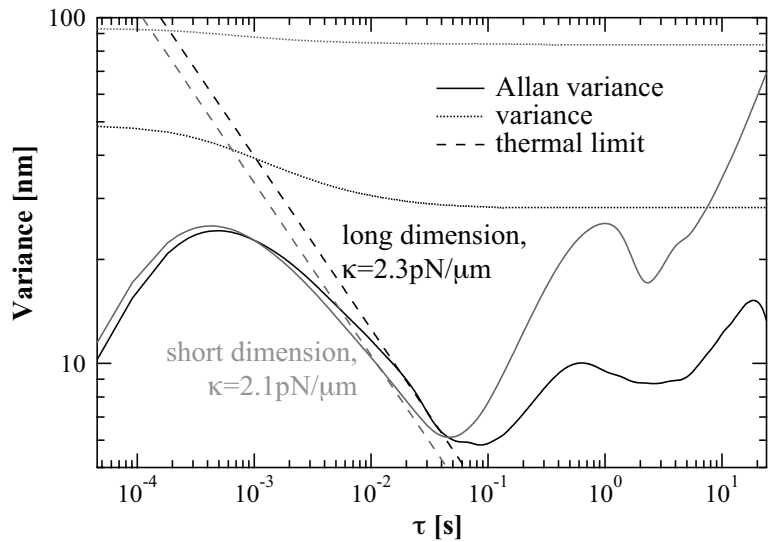}
   \end{tabular}
   \end{center}
   \caption[example] 
   { \label{fig:rods} 
Variances of optically trapped gold nanorods. The black curve is along the longest direction of the rod, $\kappa=2.3$~pN/$\mu$m,
the grey curve along the shortest direction of the rod, $\kappa=2.1$~pN/$\mu$m. The dotted lines are the corresponding normal variances, the dashed lines the thermal limits.}
   \end{figure}

So far, all figures and results presented both in the present Proceeding and in literature
~\cite{Gibson2008,Czerwinski2009} have addressed optical tweezing of micrometer-sized spherical objects. It is, however,
also possible to trap significantly smaller and non-spherical objects, such as gold nanorods. Gold nanorods as thin as 8~nm and with aspect ratios up to 5.6 have been
optically trapped~\cite{SelhuberUnkel2008}. These asymmetric nanorods align inside the trap with their
longest direction along the electrical field vector of the trapping laser and the spring constant correlates directly with the
 polarizability of the rod. Figure~\ref{fig:rods} shows the Allan variance calculated from an experiment
where an individual gold nanorod (long axis $\left(63.8\pm7.4\right)$~nm, short axis $\left(37.3\pm5.0\right)$~nm) was optically trapped. The black trace is the Allan
variance along the longest dimension of the rod, the grey trace along the shortest dimension of the rod.
The thermal limit for the shortest dimension is lower as we consider here $r$ to be the short axis (Equation~(\ref{eq:thlimitonetrap}). Figure~\ref{fig:rods} also points out that for long measurement times, the Allan variance is smaller for a stronger trapping stiffness. Moreover, it shows that the optimal measurement interval is shorter for lower thermal limits rather than for the trap stiffness.
The optimal measurement intervals are on the order of a tenth part of a second, which is significantly shorter than the optimal measurement times for the much larger polystyrene spheres, typically on the order of seconds (Figure~\ref{fig:Allan}). At measurement times longer than the absolute minimum, there are distinct peaks in the
Allan variances of both directions. As the measurement chamber was as narrow as 30~mm $\times$ 5~mm, it might be possible that those peaks originate from the geometry of the chamber~\cite{Czerwinski2009}.  The dotted lines denote the normal variances of the plots, which, also here, do not reveal any information about, e.g., optimal measurement time or the impact of drift. 


\section*{CONCLUSION}
We presented Allan variance analysis as an excellent tool to quantify noise in experiments where micro- and nanometer-sized particles were optically trapped.
Furthermore, we pointed at our proper software solutions that can easily be used for on-the-fly analyses to determine important measurement parameters, as for instance the optimal
measurement time for calibrations, or the optimal number of acquired data points for a particular experimental setting.

Fourier analysis can conveniently be used to perform
calibration procedures. Furthermore, its use to pinpoint high-frequency noise is excellent. However, for low-frequency noise which is inherently present in
every experiment, Allan variance analysis is superior. This method has been used to quantify the 
noise contribution from photodiode-based detection systems, the influence of the noise on chamber
stability and geometry~\cite{Czerwinski2009}, and, as reviewed here, the noise contribution from the piezo stage.
Allan variance analysis was mostly used on positional time series from spherical, micron-sized polystyrene spheres,
but it can be utilized even for nanometer-sized gold rods. Here, the overall
noise contribution was asymmetric, strongly correlated to the alignment of the nanorods inside the trap.

We think that Allan variance analysis, complementary and additional to Fourier analysis, allows for the development of a common standard in research that accesses optical tweezers, as it contains the possibility to compare noise and drift in different experiments, settings, setups, and even laboratories.
\section*{ACKNOWLEDGMENTS}       
 
We acknowledge discussions with M. Andersson, P. M. Bendix, and K. Berg-S\o rensen. We thank C. S\"onnichsen and I. Zins for providing the gold nanorods. The study was financially supported by a University of Copenhagen Excellence Grant. C.S.U. is supported by the German Academy of Sciences Leopoldina through grant BMBF-LPD 9901/8-164.

\bibliographystyle{plain}   

\begin{thebibliography}{10}

\bibitem{Moffitt2008}
J.~Moffitt, Y.~Chemla, S.~Smith, and C.~Bustamante, ``Recent advances in
  optical tweezers,'' {\em Ann. Rev. Biochem.}~{\bf 77}, pp.~205--228, 2008.

\bibitem{Ashkin1987}
A.~Ashkin, J.~Dziedzic, and T.~Yamane, ``Optical trapping and manipulation of
  single cells using infrared laser beams,'' {\em Nature}~{\bf 330},
  pp.~769--771, 1987.

\bibitem{Rasmussen2008}
M.~Rasmussen, L.~B.~Oddershede, and H.~Siegumfeldt, ``Optical tweezers cause
  physiological damage to {E}. coli and listeria bacteria,'' {\em Appl. Environ. Microbiol.}~{\bf 74}, pp.~2441--2446, 2008.

\bibitem{Tolic2004}
I.~M. Toli\c{c}-N{\o}rrelykke, E.-L. Munteanu, G.~Thon, L.~B. Oddershede, and
  K.~Berg-S{\o}rensen, ``Anomalous diffusion in living yeast cells,'' {\em Phys.
  Rev. Lett.}~{\bf 93}, p.~078102-, 2004.

\bibitem{Hansen2005}
P.~M.~Hansen, V.~Bhatia, N.~Harrit, and L.~B.~Oddershede, ``Expanding the optical
  trapping range of gold nanoparticles,'' {\em Nano Lett.}~{\bf 5},
  pp.~1937--1942, 2005.

\bibitem{SelhuberUnkel2008}
C.~Selhuber-Unkel, I.~Zins, O.~Schubert, C.~S{\"o}nnichsen, and L.~B.
  Oddershede, ``Quantitative optical trapping of single gold nanorods,'' {\em
  Nano Lett.}~{\bf 8}, pp.~2998--3003, 2008.

\bibitem{Bosanac2008}
L.~Bosanac, T.~Aabo, P.~M.~Bendix, and L.~B.~Oddershede, ``Efficient optical trapping
  and visualization of silver nanoparticles,'' {\em Nano Lett.}~{\bf 8},
  pp.~1486--1491, 2008.

\bibitem{Jauffred2008}
L.~Jauffred, A.~C. Richardson, and L.~B. Oddershede, ``Three-dimensional
  optical control of individual quantum dots,'' {\em Nano Lett.}~{\bf 8},
  pp.~3376--3380, 2008.

\bibitem{Abbondanzieri2005}
E.~Abbondanzieri, W.~Greenleaf, J.~Shaevitz, R.~Landick, and S.~Block, ``Direct
  observation of base-pair stepping by {RNA} polymerase,'' {\em Nature}~{\bf
  438}, pp.~460--465, 2005.

\bibitem{Carter2009}
A.~R. Carter, Y.~Seol, and T.~T. Perkins, ``Precision surface-coupled
  optical-trapping assay with one-basepair resolution,'' {\em Biophys. J.}~{\bf
  96}, pp.~2926--2934, 2009.

\bibitem{Gittes1998}
F.~Gittes and C.~Schmidt, ``Signals and noise in micromechanical
  measurements,'' {\em Methods Cell. Biol.}~{\bf 55}, pp.~129--156, 1998.

\bibitem{Klein2007}
M.~Klein, M.~Andersson, O.~Axner, and E.~Fallman, ``Dual-trap technique for
  reduction of low-frequency noise in force measuring optical tweezers,'' {\em
  Appl. Opt.}~{\bf 46}, pp.~405--412, 2007.

\bibitem{Gibson2008}
G.~M. Gibson, J.~Leach, S.~Keen, A.~J. Wright, and M.~J. Padgett, ``Measuring
  the accuracy of particle position and force in optical tweezers using
  high-speed video microscopy,'' {\em Opt. Express}~{\bf 16}, pp.~14561--14570,
  2008.

\bibitem{Czerwinski2009}
F.~Czerwinski, A.~C.~Richardson, and L.~B.~Oddershede, ``Quantifying noise in optical
  tweezers by {A}llan variance,'' {\em Opt. Express}~{\bf 17}, pp.~13255--13269, 2009.

\bibitem{Czerwinski2009a}
F.~Czerwinski, ``Bead{F}luct v1.0,'' {\em \textsc{Matlab}Central}~{\bf 24196}, 2009, \\ \textit{http://www.mathworks.com/matlabcentral/fileexchange/24196}.

\bibitem{Reihani2007}
S.~N.~S. Reihani and L.~B. Oddershede, ``Optimizing immersion media refractive
  index improves optical trapping by compensating spherical aberrations,'' {\em
  Opt. Lett.}~{\bf 32}, pp.~1998--2000, 2007.

\bibitem{Czerwinski2009b}
F.~Czerwinski and L.~B.~Oddershede, ``Reliable data-streaming software for
  photodiode readout in \textsc{Labview},'' {\em in prep.} , 2009.

\bibitem{Oddershede2001}
L.~B. Oddershede, S.~Grego, S.~N{\o}rrelykke, and K.~Berg-S{\o}rensen,
  ``Optical tweezers: probing biological surfaces,'' {\em Probe Microsc.}~{\bf
  2}, pp.~129--137, 2001.

\bibitem{BergSorensen2004}
K.~Berg-S{\o}rensen and H.~Flyvbjerg, ``Power spectrum analysis for optical
  tweezers,'' {\em Rev. Sci. Instrum.}~{\bf 75}, pp.~594--612, 2004.

\bibitem{Hansen2006b}
P.~M.~Hansen, I.~Toli\c{c}-N\o{}rrelykke, H.~Flyvbjerg, and K.~Berg-S\o{}rensen,
  ``tweezercalib 2.1: Faster version of \textsc{Matlab} package for precise calibration
  of optical tweezers'' {\em Comput. Phys. Commun.}~{\bf 175},
  pp.~572--573, 2006.

\bibitem{Allan1966}
D.~W. Allan, ``Statistics of atomic frequency standards," {\em Proc. IEEE}
  \textbf{54}, 221--230, 1966.

\bibitem{Czerwinski2008}
F.~Czerwinski, ``allan v1.71,'' {\em \textsc{Matlab}Central}~{\bf 21727}, 2008, \\ \textit{http://www.mathworks.com/matlabcentral/fileexchange/21727}.

\bibitem{BergSorensen2003}
K.~Berg-S\o{}rensen, L.~B.~Oddershede, E.~Florin, and H.~Flyvbjerg, ``Unintended
  filtering in a typical photodiode detection system for optical tweezers,''
  {\em J. Appl. Phys.}~{\bf 93}, pp.~3167--3176, 2003.

\end{thebibliography}

\end{document}